\documentclass[
reprint,
superscriptaddress,
aps,
prl,
]{revtex4-2}

\usepackage{graphicx}
\usepackage{dcolumn}
\usepackage{bm}
\usepackage[utf8]{inputenc}
\usepackage{amsmath, amsthm, amssymb, amsfonts}
\usepackage{accents}
\usepackage[usenames,dvipsnames]{xcolor}
\usepackage[caption=false]{subfig}
\usepackage{tikz}
\usepackage{array}
\usepackage{dsfont}
\usepackage{mathtools}
\usepackage{soul,xcolor}
\usepackage[colorlinks=true,linkcolor=Blue,citecolor=Blue,urlcolor=Blue]{hyperref}
\usepackage[capitalise]{cleveref}

\usepackage[version=4]{mhchem}
\usepackage{siunitx}

\setstcolor{magenta}

\begin{document}

\title{Random heterogeneity outperforms design in network synchronization}

\author{Yuanzhao Zhang}
\affiliation{Department of Physics and Astronomy, Northwestern University, Evanston, Illinois 60208, USA}
\affiliation{Center for Applied Mathematics, Cornell University, Ithaca, New York 14853, USA}

\author{Jorge L. Ocampo-Espindola}
\affiliation{Department of Chemistry, Saint Louis University, St. Louis, Missouri 63103, USA}

\author{István Z. Kiss}
\affiliation{Department of Chemistry, Saint Louis University, St. Louis, Missouri 63103, USA}

\author{Adilson E. Motter}
\affiliation{Department of Physics and Astronomy, Northwestern University, Evanston, Illinois 60208, USA}
\affiliation{Northwestern Institute on Complex Systems, Northwestern University, Evanston, Illinois 60208, USA}

\begin{abstract}
A widely held assumption on network dynamics is that similar components are more likely to exhibit similar behavior than dissimilar ones and that generic differences among them are necessarily detrimental to synchronization. Here, we show that this assumption does not generally hold in oscillator networks when communication delays are present. We demonstrate, in particular, that random parameter heterogeneity among oscillators can consistently rescue the system from losing synchrony. This finding is supported by electrochemical-oscillator experiments performed on a multi-electrode array network. Remarkably, at intermediate levels of heterogeneity, random mismatches are more effective in promoting synchronization than parameter assignments specifically designed to facilitate identical synchronization. Our results suggest that, rather than being eliminated or ignored, intrinsic disorder in technological and biological systems can be harnessed to help maintain coherence required for function.

\vspace{3mm}
\noindent DOI: \href{https://doi.org/10.1073/pnas.2024299118}{10.1073/pnas.2024299118} 
\end{abstract}

%\pacs{05.45.Xt, 89.75.Fb}

\maketitle

\section*{Introduction}

Heterogeneity among interacting components is usually seen as detrimental to the emergence of uniform dynamics in networks, including consensus \cite{olfati2007consensus,manrique2018individual} and synchronization \cite{strogatz2000kuramoto,arenas2008synchronization}.
For networks of coupled oscillators, the assumption has been that global synchronization would be hindered by parameter mismatches among oscillators. 
This assumption, which is observed to hold for Kuramoto models \cite{strogatz2000kuramoto}, remains under-substantiated for more general classes of oscillator networks \cite{wiley2006size,li2010consensus,o2015synchronization}, especially those studied using the master stability function formalism and its variants \cite{pecora1998master,nishikawa2006synchronization,choe2010controlling,tang2019master}.
In the relatively few theoretical studies that have explicitly considered oscillator heterogeneity beyond the context of Kuramoto models, the focus has been on small parameter mismatches and the persistence of synchronization among nearly-identical oscillators \cite{restrepo2004spatial,sun2009master,pereira2014towards,acharyya2015synchronization,sorrentino2016approximate}.
These results all conform to the perception that disorder, in the form of random oscillator heterogeneity, is undesirable for synchronization.

Yet, a few exceptions to this perception exist in the literature. 
In particular, it has been shown that disorder can sometimes enhance synchronization and/or spatiotemporal order in arrays of driven dissipative pendulums \cite{braiman1995taming,braiman1995disorder,brandt2006synchronization}.
For example, for coupled oscillators in a chaotic regime, heterogeneity can suppress chaos, giving rise to more regular patterns \cite{braiman1995taming}.
In these studies, and numerous subsequent ones \cite{tessone2006diversity,tessone2007theory,montaseri2016diversity}, disorder does not stabilize the system around an original synchronization orbit, but instead generates new behavior that is qualitatively different.
However, it has been recently realized that certain oscillator heterogeneities can stabilize a synchronization orbit of the homogeneous system \cite{nishikawa2016symmetric,zhang2017asymmetry,zhang2017nonlinearity}.
In these studies, the heterogeneity is purposively designed to preserve at least one common orbit among the nonidentical oscillators, which may not always be practical to achieve in applications.

Here, we show that oftentimes {\it random} differences among individual oscillators can {\it consistently} stabilize the dynamics around an otherwise unstable synchronization orbit of the homogeneous system.
We demonstrate the phenomenon for random heterogeneity in delay-coupled Stuart-Landau oscillators, which is a canonical model for limit-cycle oscillations close to a Hopf bifurcation. 
Stuart-Landau oscillators have been used to describe numerous processes, ranging from electrochemical reaction oscillations \cite{zou2015restoration} to plant circadian rhythms \cite{fukuda2007synchronization}.
Importantly, we establish that, to preserve system-level coherence, random heterogeneity can be more effective than heterogeneity purposely designed to optimize the stability of identical synchronization. 
To support our theoretical and numerical results, we performed experiments using coupled electrochemical oscillators.
The experimental results confirm our predictions and further demonstrate the effect for systems not in the close vicinity of a Hopf bifurcation.
These findings are expected to have implications for a broad class of natural and engineered systems, whose functions depend on the synchronization of heterogeneous components.

\section*{Results}

\subsection*{Modeling the Dynamics of Heterogeneous Oscillators}

\begin{figure*}[t]
\centering
\includegraphics[width=1\linewidth]{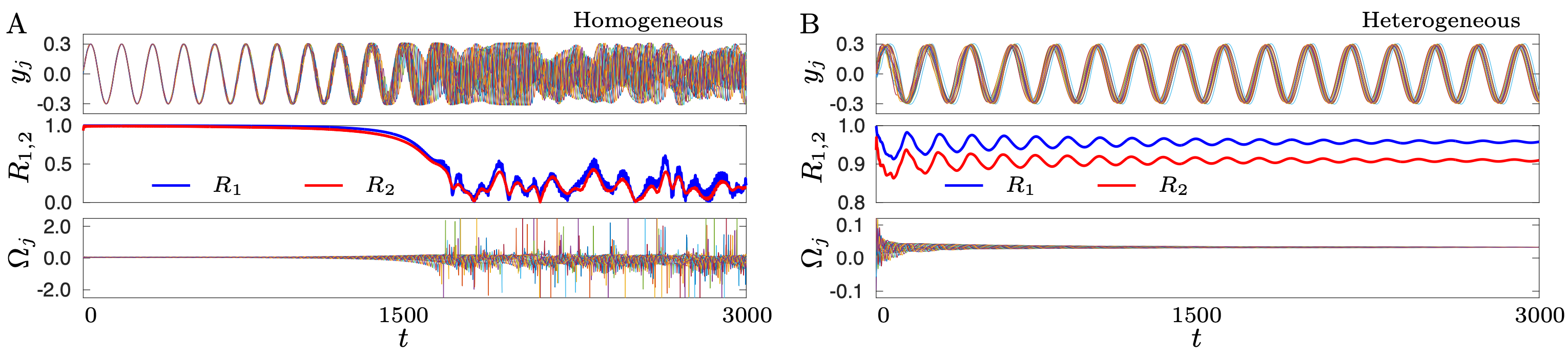}
\vspace{-6mm}
\caption{Impact of oscillator heterogeneity on the synchronization dynamics of Stuart-Landau oscillators.
The different panels show the time evolution of the imaginary components $y_j$ (top), order parameters $R_1$ and $R_2$ (middle), and angular velocities $\Omega_j = \dot{\psi}_j$ (bottom), for trajectories in a $18$-node ring network initialized close to the identical synchronization state.
({\it A}) Homogeneous system for the parameters defined in the text.
({\it B}) Heterogeneous system for the same parameters, except for base angular velocities $\{\omega_j\}$, which are drawn from a Gaussian distribution with standard deviation $\sigma=0.1$.
The trajectories show that synchronization is unstable in the homogeneous system but becomes stable in the heterogeneous one.
}
\label{fig:1}
\end{figure*}

We consider a network of $N$ delay-coupled nonidentical Stuart-Landau oscillators, whose dynamics are governed by
\begin{equation}
  \dot{z}_j(t) = f_j(z_j(t)) + \frac{K}{d_j} \sum_{k=1}^N A_{jk} \left[ z_k(t-\tau) - z_j(t) \right],
  \label{eq:1}
\end{equation}
\hspace{-.5mm}where $z_j=r_je^{\mathrm{i}\psi_j} = x_j + \mathrm{i}y_j$ is a complex variable representing the state of the $j$th oscillator and $K$ is the coupling strength.
The adjacency matrix $\bm{A} = \{A_{jk}\}$ encodes the network structure and $d_j = \sum_k A_{jk}$ is the indegree for oscillator $j$.
The coupling delay $\tau$ models the finite speed of signal propagation in real systems, which is often significant in biological \cite{ernst1995synchronization,bratsun2005delay,vicente2008dynamical}, physical \cite{fischer2006zero,flunkert2009bubbling,bick2017robust}, and control systems \cite{pyragas1992continuous,gu2003survey}.
The local dynamics $f_j$ are expressed in the following canonical form for systems born out of a Hopf bifurcation \cite{kuramoto2012chemical}:
\begin{equation}
  f_j(z_j) = \left[ \lambda_j + \mathrm{i}\omega_j - (1+\mathrm{i} \gamma_j)|z_j|^2 \right]z_j,
  \label{eq:2}
\end{equation}
where $\lambda_j$, $\omega_j$, and $\gamma_j$ are real parameters associated with the amplitude, base frequency, and amplitude-dependent frequency of the underlying limit-cycle oscillations.

The oscillators are identical when $\lambda_j = \lambda, \; \omega_j = \omega$, and $\gamma_j = \gamma$ for all $j$.
For identical oscillators, the identical synchronization state is defined by the limit-cycle synchronous solution
\begin{equation}
    z_j = r_0 e^{\mathrm{i} \Omega_0 t}, \quad j=1,\cdots,N.
\label{eq:3}
\end{equation}
The amplitude $r_0$ and angular velocity $\Omega_0$ can be found by solving the transcendental equations
\begin{subequations}
    \begin{align}
      r_0^2 &= \lambda + K (\cos \Phi - 1), \\
      \Omega_0 &=\omega - \gamma r_0^2 + K \sin \Phi,
    \end{align}
\label{eq:4}
\end{subequations}
\hspace{-1.5mm}where $\Phi = - \Omega_0 \tau$ \cite{choe2010controlling}. 
When random heterogeneity is introduced through one or more oscillator parameters, the identical synchronization state described by Eq.~\ref{eq:3} will, in general, no longer exist.
Nonetheless, we show that heterogeneous systems can still admit states that are synchronized in the sense of exhibiting cohesive phase and amplitude dynamics, as formalized below. 
Here, we consider synchronization in this sense and ask whether it can be stabilized by random oscillator heterogeneity.

We start by considering a homogeneous system consisting of $N=18$ identical Stuart-Landau oscillators coupled through a directed ring network ($A_{jk}=1$ if $k=j+1 \mod N$ and $A_{jk}=0$ otherwise) for $\lambda = 0.1$, $\omega = -0.28$, $\gamma = -4.42$, $K = 0.3$, and $\tau = 1.8\pi$. 
Under this parameter choice, the limit-cycle synchronous solution described by Eq.~\ref{eq:3}
%($r^2,\,\Omega$) = (0.0904, 0.0448) 
is unstable, and the system evolves into a symmetry-broken state exhibiting incoherent chaotic dynamics.
As an example of a heterogeneous system, we consider the same network with the base angular velocity $\omega_j$ drawn from a Gaussian distribution with mean $\omega = -0.28$ (as in the homogeneous system) and standard deviation $\sigma = 0.1$.
For additional details on the numerical procedure, see {\it Materials and Methods}.

In \cref{fig:1}, we show typical trajectories, order parameters, and angular velocities for the homogeneous and heterogeneous systems. 
Here, two order parameters are introduced to measure the cohesiveness of the dynamics: the phase order parameter $R_1 = \big\lvert \sum_j e^{\mathrm{i} \psi_j}/N \big\rvert$, which is the one typically used in the study of Kuramoto oscillators; the phase-amplitude order parameter $R_2 = \frac{1}{\text{max}(r_j)} \big\lvert \sum_j r_j e^{\mathrm{i} \psi_j}/N \big\rvert$, which measures the coherence in both phases and amplitudes.
It follows that both $R_1$ and $R_2$ are constant for frequency-synchronized states.
For the trajectories shown, the two systems were initialized close to the limit-cycle synchronous state.
The homogeneous system loses synchrony at $t\approx1500$ and transitions to an incoherent state with both $R_1$ and $R_2$ fluctuating around 0.2. 
Remarkably, despite having different base frequencies, oscillators in the heterogeneous system converge to a stable cohesive state with large order parameters ($R_1>R_2>0.9$) and identical angular velocities. 
That is, the oscillators are not only approximately synchronized in phase and amplitude---they are also exactly synchronized in frequency (i.e., phase-locked).
For an animation of this phenomenon in larger systems, see Movie S1. 
It is worth noting that the same phenomenon is also observed when the coupling function is nonlinear and/or when the coupling delay is link-dependent, which we demonstrate in {\it SI Text} and Figs.~S1--S2.

\subsection*{Synchronization States and Stability Conditions}
To gain theoretical understanding of the effect shown in \cref{fig:1} and its prevalence, we characterize the synchronization states of interest in the presence of heterogeneity and derive analytical conditions for their stability.
These results are established for delay-coupled Stuart-Landau oscillators with arbitrary heterogeneity.

\begin{figure}[t]
\centering
\subfloat[]{
\includegraphics[width=1\columnwidth]{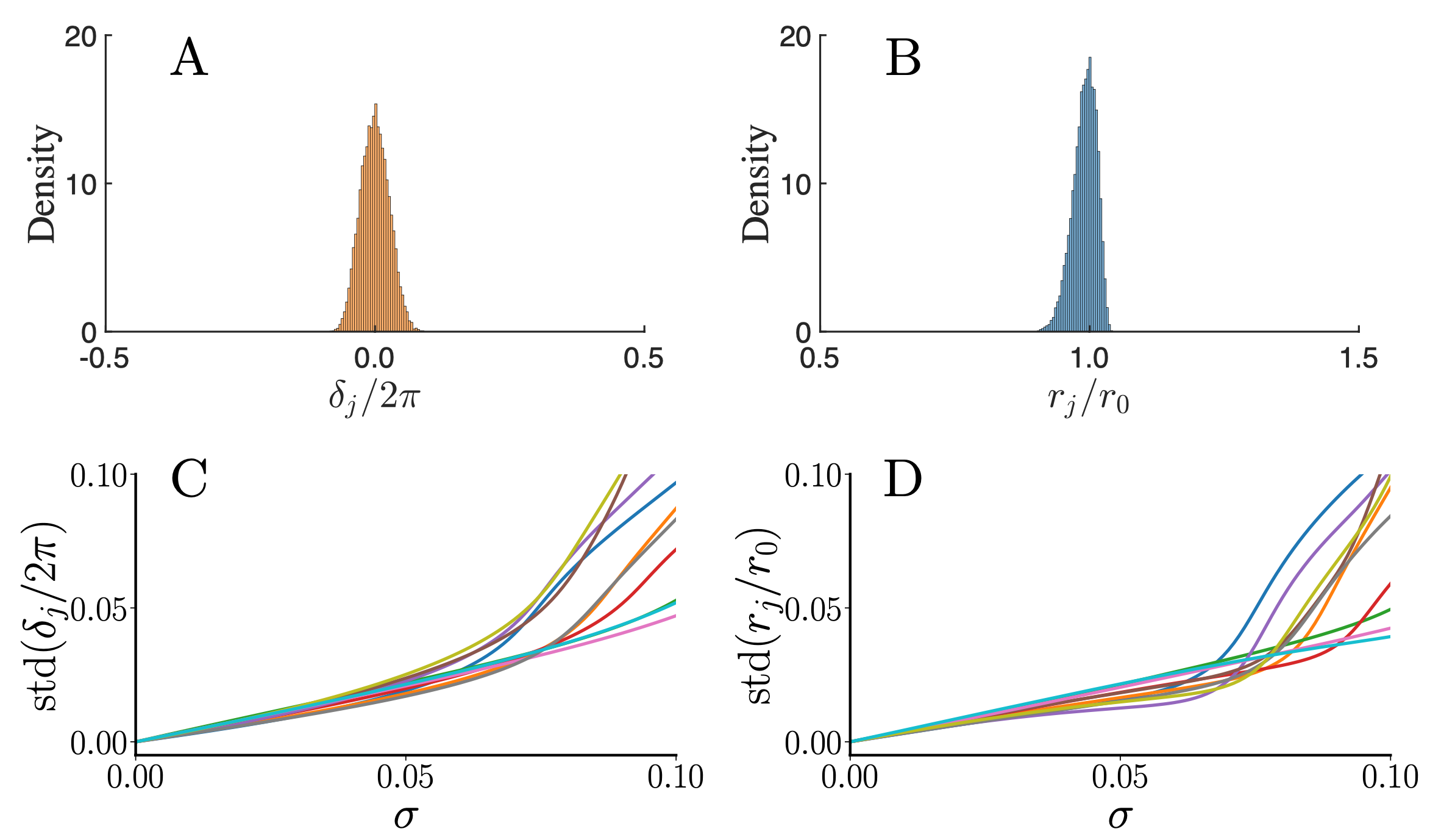}
}
\vspace{-6mm}
\caption{Normalized phase lags $\delta_j$ and amplitudes $r_j$ determined by Eqs.~\ref{eq:s1}a and \ref{eq:s1}b for heterogeneous $\omega_j$. 
({\it A--B}) Distribution of $\delta_j/2\pi$ ({\it A}) and $r_j/r_0$ ({\it B}) for $\sigma = 0.06$ estimated from $1000$ random realizations of heterogeneity. 
({\it C--D}) Standard deviation of $\delta_j/2\pi$ ({\it C}) and $r_j/r_0$ ({\it D}) as functions of $\sigma$, where each curve corresponds to an independent realization of heterogeneity scaled by $\sigma$ ($10$ realizations in total). 
The network and the other parameters are the same as in \cref{fig:1}.
}
\label{fig:2}
\end{figure}

Inspired by the fact that the heterogeneous system in \cref{fig:1}B settles into a frequency-synchronized state, for which the frequencies of all oscillators are equal and the phase differences and amplitudes remain constant, we employ the following ansatz:
\begin{equation}
    z_j = r_j e^{\mathrm{i}(\Omega t + \delta_j)},
\end{equation}
where oscillator $j$ has amplitude $r_j$ and phase lag $\delta_j$, both of which are assumed to be constant, and all oscillators share the same angular velocity $\Omega$. 
Substituting this ansatz into Eq.~\ref{eq:1}, we obtain $2N$ nonlinear algebraic equations with $2N$ unknowns:
\begin{subequations}
    \begin{align}
      r_j^2 &= \lambda_j + \frac{K}{d_j} \sum_{k=1}^N A_{jk} \frac{r_k}{r_j} \cos \Phi_{jk} - K, \\
      \Omega &= \omega_j - \gamma_j r_j^2 + \frac{K}{d_j} \sum_{k=1}^N A_{jk} \frac{r_k}{r_j} \sin \Phi_{jk},
    \end{align}
    \label{eq:s1}
\end{subequations}
\hspace*{-1mm}for $j=1,\cdots,N$, where $\Phi_{jk} = \delta_k - \delta_j - \Omega\tau$. 
Taking $\delta_1=0$, which can be done without loss of generality, the solution of Eqs.~\ref{eq:s1}a and \ref{eq:s1}b determines $\delta_2,\cdots,\delta_N,r_1,\cdots,r_N$, and $\Omega$. 
As shown in \cref{fig:2}, when parameter heterogeneity is not too large, this gives us frequency-synchronized states that are close to the identical synchronization state of the homogeneous system given by Eqs.~\ref{eq:3} and \ref{eq:4}.

We can analyze the stability of the frequency-synchronized states through the variational equation that governs the evolution of small deviations $\delta r_j(t)$ and $\delta \psi_j(t)$ from those states. 
Taking $z_j(t) = r_j[1+\delta r_j(t)]e^{\mathrm{i}[\Omega t + \delta_j + \delta\psi_j(t)]}$, $\bm{\eta}_j = (\delta r_j, \delta \psi_j)^\intercal$, and $\bm{\eta} = (\bm{\eta}_1^\intercal,\cdots,\bm{\eta}_N^\intercal)^\intercal$, the variational equation can be written as
\begin{equation}
    \dot{\bm{\eta}}(t) = \oplus\left( \bm{J}_j - K\bm{P}_j \right) \bm{\eta}(t) + K (\bm{D} \bm{A} \otimes \bm{R}_{jk}) \bm{\eta}(t-\tau),
    \label{eq:s2}
\end{equation}
where $\bm{J}_j = \big(\begin{smallmatrix} -2r_j^2 & 0 \\ -2\gamma_jr_j^2 & 0 \end{smallmatrix}\big)$, 
$\bm{D} = \text{diag}(\frac{1}{d_1},\cdots,\frac{1}{d_N})$,
$\bm{R}_{jk} = \frac{r_k}{r_j}\big(\begin{smallmatrix} \cos\Phi_{jk} & -\sin\Phi_{jk} \\ \sin\Phi_{jk} & \cos\Phi_{jk} \end{smallmatrix}\big)$, 
$\bm{P}_j = \frac{1}{d_j}\sum_k A_{jk} \bm{R}_{jk}$, and
$\oplus (\bm{J}_j - K\bm{P}_j) = \text{diag}(\bm{J}_1 - K\bm{P}_1,\cdots,\bm{J}_N - K\bm{P}_N)$.
Here, $\bm{D}\bm{A} \otimes \bm{R}_{jk}$ is a $2N \times 2N$ matrix obtained by replacing each entry $A_{jk}/d_j$ in $\bm{D}\bm{A}$ with a $2\times2$ block given by $A_{jk}\bm{R}_{jk}/d_j$.

Since all the matrices in Eq.~\ref{eq:s2} are time independent, 
we can assume $\bm{\eta}(t) = \bm{\eta}(0)e^{v_\ell t}$ and reduce the stability calculation to determining the exponents $v_\ell$ that solve the following characteristic equation:
\begin{equation}
    \text{det}\{\oplus\left( \bm{J}_j - K\bm{P}_j \right) + K (\bm{D} \bm{A} \otimes \bm{R}_{jk}) e^{-v_\ell \tau} - v_\ell \mathds{1}_{2N}\}=0,
    \label{eq:s3}
\end{equation}
where $\ell$ indexes the solutions.
The Lyapunov exponents of Eq.~\ref{eq:s2} are given by the real parts of $v_\ell$. 
One Lyapunov exponent (referred to as $\text{Re}(v_0)$) is identically null and corresponds to the perturbation mode parallel to the frequency-synchronization manifold (in which the phases of all oscillators are subject to the same perturbation).
The other Lyapunov exponents correspond to perturbation modes transverse to the synchronization manifold.
The stability of the frequency-synchronization state is determined by the sign of the maximum transverse Lyapunov exponent (MTLE), which is given by $\Lambda = \max_{\ell\neq 0} \{\text{Re}(v_\ell)\}$.

\subsection*{Disorder Consistently Promotes Synchronization}
We now examine systematically the phenomenon of synchronization induced by random heterogeneity.
In particular, we address key questions underlying its prevalence. 
For example, what is the effect of the magnitude of parameter mismatches?  
Do the results change significantly depending on which parameters are made heterogeneous?
Most importantly, can different realizations of random heterogeneity consistently induce synchronization?

\begin{figure*}[t]
\centering
\includegraphics[width=1\linewidth]{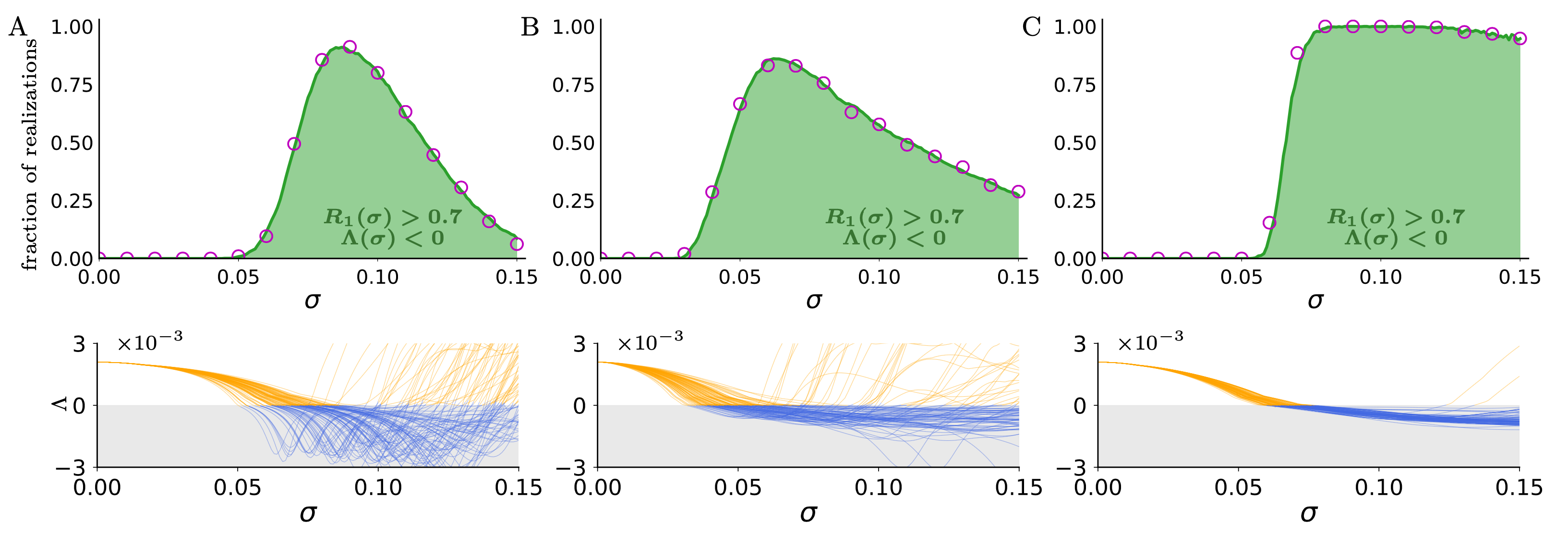}
\vspace{-5mm}
\caption{
Statistics on the synchronizing effect of random oscillator heterogeneity for systems with ({\it A}) nonidentical $\omega_j$, ({\it B}) nonidentical $\lambda_j$, and ({\it C}) nonidentical $\gamma_j$.
For each upper panel, we generate $1000$ realizations of random heterogeneity in the corresponding parameter and then, for each realization, calculate $R_1(\sigma)$ and $\Lambda(\sigma)$ of the frequency-synchronized state as $\sigma$ is increased from $0$ to $0.15$.
The filled green curves show the percentage of realizations that successfully stabilize a frequency-synchronized state with order parameter $R_1>0.7$, which are validated by direct simulations shown as purple circles.
To visualize the differences between different realizations and their characteristics as an ensemble, each lower panel shows $\Lambda(\sigma)$ for $100$ independent realizations of random heterogeneity.
The stable portions are highlighted in blue.
The network and other parameters are the same as in \cref{fig:1}.
}
\label{fig:3}
\end{figure*}

In \cref{fig:3}, we start with the same homogeneous system as in \cref{fig:1}A and introduce heterogeneity in $\{\omega_j\}$, $\{\lambda_j\}$, and $r_0^2\{\gamma_j\}$, respectively. 
(Here, the factor $r_0^2$ is introduced to scale $\sigma$ for $\gamma_j$, because the influence of $\gamma_j$ in Eq.~\ref{eq:2} is scaled by the square of the oscillation amplitude. The constant $r_0$ can be found by solving Eqs.~\ref{eq:4}a and \ref{eq:4}b for the corresponding homogeneous system.)
In all cases, the standard deviation is $\sigma$ and the mean is taken to be the same as the corresponding parameter in the homogeneous system.
For each realization of heterogeneity, as $\sigma$ increases from zero, the identical synchronization state progressively changes into a phase-locked state with large order parameters.
The stability of this state is measured by $\Lambda(\sigma)$, which we obtain by solving Eq.~\ref{eq:s3} for each realization of heterogeneity.
The filled green curves in the upper panels show the probability that synchronization is stabilized by random heterogeneity in each parameter. 
These results are verified by direct simulations of Eqs.~\ref{eq:1} and \ref{eq:2} for various $\sigma$, shown as purple circles.
In the lower panels, we plot $\Lambda(\sigma)$ for a representative subset of realizations of heterogeneity in each parameter, visualizing their impact on stability as an ensemble.

One can see from \cref{fig:3} that there is always a sweet spot of optimal heterogeneity at an intermediate value of $\sigma$.
Around that sweet spot, the green curves stay very close to $1$, indicating that intermediate levels of heterogeneity can consistently induce synchronization, largely independent of its particular realization.
It is interesting to note from the lower panels that small heterogeneity always improves stability under the conditions considered, as reflected in the monotonic decrease of $\Lambda(\sigma)$ for small $\sigma$.
Disorder can also consistently stabilize synchronization when all three parameters are allowed to be heterogeneous, as demonstrated in {\it SI Text} and Fig.~S3.
Furthermore, we verified that the same effect can be observed for a wide range of network sizes and different network structures (see {\it SI Text} and Figs.~S4--S6).

\subsection*{Disorder Can Be Better Than Design}

\begin{figure}[tb]
\centering
\includegraphics[width=1\columnwidth]{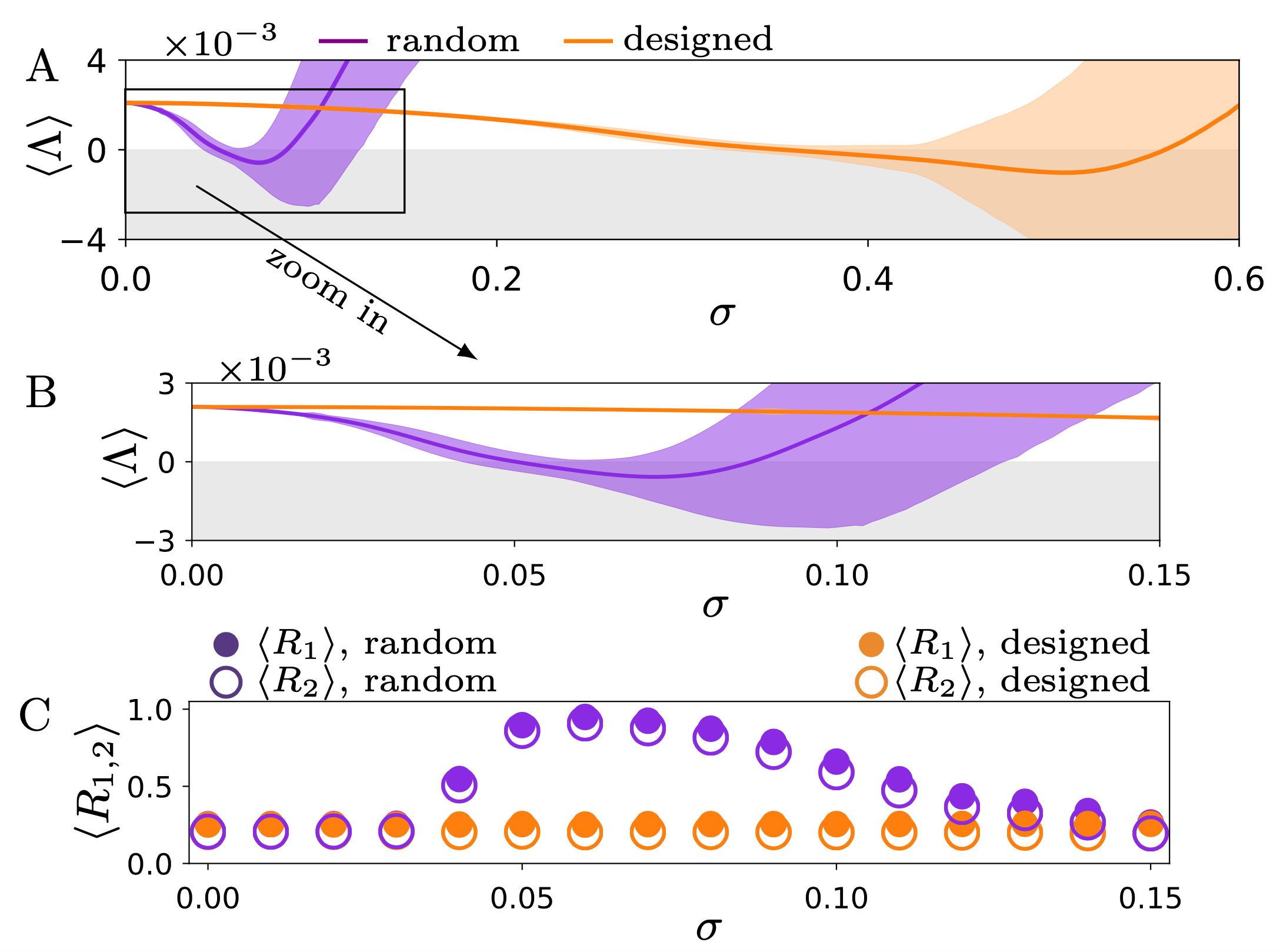}
\vspace{-5mm}
\caption{Comparing systems with random and designed heterogeneity in $\{\omega_j\}$ and $\{\gamma_j\}$, where the latter preserves a common orbit among heterogeneous oscillators. 
({\it A}) Average MTLE of systems with random heterogeneity (purple line) and designed heterogeneity (orange line). The shades indicate the standard deviation among 1000 independent realizations. Designed heterogeneity stabilizes synchronization when $\sigma$ is large, but fails to do so for intermediate $\sigma$, where random heterogeneity is effective.
({\it B}) Magnification of the marked portion of {\it A}, highlighting the effectiveness of random heterogeneity of intermediate magnitude.
({\it C}) Average order parameters of systems with random and designed heterogeneities. Each data point is averaged over 1000 independent realizations of heterogeneity and also averaged over time for steady states that are not frequency synchronized. 
The network and other parameters are the same as in \cref{fig:1}.
}
\label{fig:4}
\end{figure}

It is important to compare the effect of random and nonrandom heterogeneities.
When the heterogeneity is purposively designed, Stuart-Landau oscillators can synchronize identically (i.e., all phase differences are identically zero and all amplitudes are equal) even though they are nonidentical. 
This is most easily seen from Eqs.~\ref{eq:4}a and \ref{eq:4}b, whose solution remains invariant under the transformation $\omega \rightarrow \omega + h$, $\gamma \rightarrow \gamma + h/r_0^2$ for any $h \in \mathbb{R}$. 
Thus, any given Stuart-Landau oscillator belongs to a continuous family of nonidentical Stuart-Landau oscillators parameterized by $h$, within which the oscillators can synchronize identically with each other.
Moreover, as shown in Ref.~\cite{zhang2017nonlinearity}, mixing different oscillators from the same family can stabilize identical synchronization that would otherwise be unstable.

By designing heterogeneity to preserve identical synchronization, can we do better than by relying on random heterogeneity?
Once again we start with the homogeneous system studied in \cref{fig:1,fig:3}.
The oscillators are then made heterogeneous by sampling from the identically synchronizable family, with $h$ drawn from a Gaussian distribution.
More concretely, $\omega_j = \omega + h_j$ and $\gamma_j = \gamma + h_j/r_0^2$, where $\{h_j\}$ has standard deviation $\sigma$ and mean zero.
This can be seen as a special subset of oscillators with random heterogeneity in both parameters $\{\omega_j\}$ and $\{\gamma_j\}$, the crucial difference being that $\omega_j - \omega$ and $\gamma_j - \gamma$ are not independent for the designed heterogeneity.

In \cref{fig:4}, we compare the ensemble average MTLE and order parameters between systems with random heterogeneity and systems with designed heterogeneity on the same parameters.
Consistent with \cref{fig:3}, random heterogeneity is most effective for  intermediate magnitudes $\sigma$, ranging from $0.05$ to $0.1$. 
On the other hand, designed heterogeneity is effective for much larger $\sigma$, from about 0.4 to 0.6, which may be interpreted as a consequence of the identical synchronization solution being preserved in this case. 
Remarkably, no system with designed heterogeneity is stable within the range for which random heterogeneity is effective. 
This implies that at intermediate magnitude, random heterogeneity can outperform heterogeneity specifically designed to preserve identical synchronization.

\subsection*{Insight From a Minimal System}

\begin{figure*}[t]
\centering
\includegraphics[width=1\linewidth]{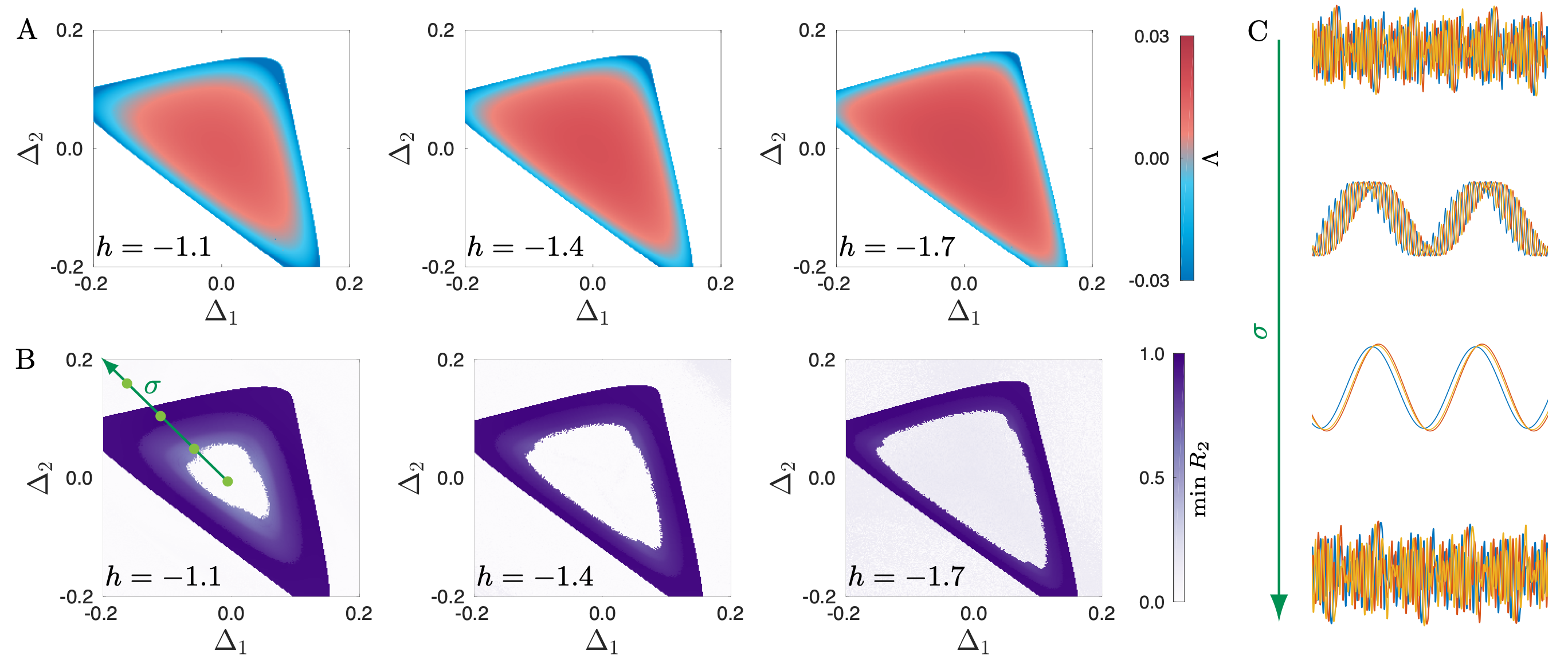}
\vspace{-5mm}
\caption{Synchronization among three nonidentical Stuart-Landau oscillators coupled through a directed ring network. 
({\it A}) MTLE of the phase-locked state in the $\Delta_1$--$\Delta_2$ plane.  
The homogeneous system lies at the origin of each panel, and its instability increases as $h$ is changed from $-1.1$ to $-1.7$.
For small $\Delta_1$ and $\Delta_2$, heterogeneity is not strong enough to tame the instability (red regions).
At intermediate $\Delta_1$ and/or $\Delta_2$, a stability belt emerges (blue regions), demonstrating the stabilizing effect of heterogeneity.
When heterogeneity becomes too strong, however, the phase-locked solution no longer exists (blank regions).
({\it B}) Minimum of the order parameter $R_2$ over 10000 time units in a steady state. 
As one moves along the green line for increasing $\sigma$, we observe the incoherence-coherence-incoherence transitions predicted by the stability analysis. 
({\it C}) Time evolution of $y_j$ (colored by oscillator) for representative states corresponding to the parameters marked by the dots in the left panel in {\it B}.
In all panels, the other parameters are $\lambda = 0.1$, $\omega = 1$, $\gamma = 0$, $K =0.3$, and $\tau = 1.8\pi$.
}
\label{fig:5}
\end{figure*}

To gain further understanding, in \cref{fig:5} we focus on a minimal system formed by three nonidentical Stuart-Landau oscillators coupled through a directed ring network. 
The $j$th oscillator has parameters $\{\lambda_j,\omega_j,\gamma_j\} = \{\lambda,\omega+h,\gamma+(h+\Delta_j)/r_0^2\}$, with the constraint that $\sum_{j=1}^3 \Delta_j = 0$. 
The parameter $h$ is introduced to vary the synchronization stability of the homogeneous system without altering the synchronous solution.
This enables us to investigate all possible realizations of heterogeneous $\gamma_j$ for different levels of instability by sweeping the $\Delta_1$--$\Delta_2$ plane.

In Figs.~\ref{fig:5}{{\it A} and {\it B}, the origin is the only point corresponding to a homogeneous system, and the differences among oscillators increase as one moves away from the origin along the radial directions. 
Stability analysis indicates that regions of stability appear for intermediate levels of oscillator heterogeneity, as shown in \cref{fig:5}{{\it A} ($\Lambda<0$, blue belts). 
The phase-locked state is unstable for weak disorder ($\Lambda>0$, red areas) and ceases to exist for strong disorder (blank areas).
A complementary perspective is offered by direct simulations, as shown in \cref{fig:5}{{\it B}.
Because order parameters averaged over time is a poor indicator of coherence for systems with a small number of oscillators, we quantify the level of coherence using the minimum of $R_2$ over a period of 10000 time units after the initial transient. 
For zero and small heterogeneity, the three oscillators are in an incoherent state with $\min R_2 \approx 0$.  
As $\sigma$ is increased further, the oscillators first settle into an approximate synchronization state with $\min R_2$ ranging from $0.6$ to $0.9$ (light purple regions). 
The level of coherence continues to improve until it plateaus at $\min R_2 \approx 0.96$ for phase-locked states (dark purple regions), which correspond to the stable states marked by the blue belts in \cref{fig:5}{{\it A}.
Finally, once we cross the outer boundary, synchrony is lost again and the value of $\min R_2$ falls back to approximately $0$.
This incoherence-coherence-incoherence transition is illustrated in \cref{fig:5}{{\it C} with representative trajectories from each stage.
It is worth noting that even before the phase-locked state is fully stabilized, disorder can already induce approximate synchronization states with well-defined rhythms, as illustrated by the second trajectory. 

\Cref{fig:5} demonstrates two competing effects of disorder: when heterogeneity is too small, it cannot tame synchronization instability; when it is too large, it destroys the synchronization state. 
In other words, there is a trade-off between synchronizability and stability, and stable synchronization naturally emerges at intermediate levels of disorder.
Another interesting observation is that the stable belts are contiguous in all cases in \cref{fig:5}{\it A} and completely surround the unstable regions in the middle, which explains why intermediate levels of heterogeneity can consistently stabilize synchronization.
It also demonstrates that the effect is robust against increasing instability (controlled by $h$) in the homogeneous system.

\subsection*{Electrochemical Experiments}
 
A natural question at this point is whether the described phenomenon is robust and general enough to be observed in real systems.
To provide an answer, we performed experiments using chemical oscillators based on the electrochemical dissolution of nickel in sulfuric acidic media \cite{zou2015restoration}.
The experimental apparatus consists of a counter electrode, a reference electrode, a potentiostat, and $N$ nickel wires submerged in the same sulfuric acidic media, each attached to a resistor (\cref{fig:6}{{\it A}}). 
At constant circuit potential ($V_0 = 1.24$ V relative to the reference electrode) and with the resistance of resistors set to $\xi_j = 1.06$ kohm, the dissolution rate of each nickel wire, measured as its current, exhibits periodic oscillations \cite{Kiss:1999wt}. 
The oscillatory dynamics originate from a Hopf bifurcation at $V_0 = 1.07$ V.
Compared to the circuit potential used in some previous studies \cite{bick2017robust}, the system here is farther away from the bifurcation point.
When the wires are placed sufficiently far from each other, the current oscillations do not show noticeable synchronization, confirming that the interactions through the solution are negligible. 
Coupling among the wires can be introduced through external feedback \cite{zou2015restoration,bick2017robust}, in which the circuit potentials of the wires $V_j(t)$ are set based on the measured currents $I_j(t)$ as
\begin{equation}
   V_j(t) = V_0 + \frac{K}{d_j} \sum_{k=1}^{N} A_{jk} \left[I_k(t-\tau) - I_j(t)\right] %\quad j=1,\cdots,N,
\label{eq:exp}
\end{equation}
for $j=1,\cdots,N$, where $K$ and $\tau$ are the experimental coupling strength and delay, respectively. 
Here, we investigate $N=16$ wires with oscillatory currents arranged in an undirected $4$-by-$4$ lattice network with periodic boundary conditions, which can be seen as a 2-dimensional variant of the ring networks considered above.
For additional details on the experimental setup and procedure, see {\it Materials and Methods}.

\begin{figure*}[t]
\centering
\includegraphics[width=1.6\columnwidth]{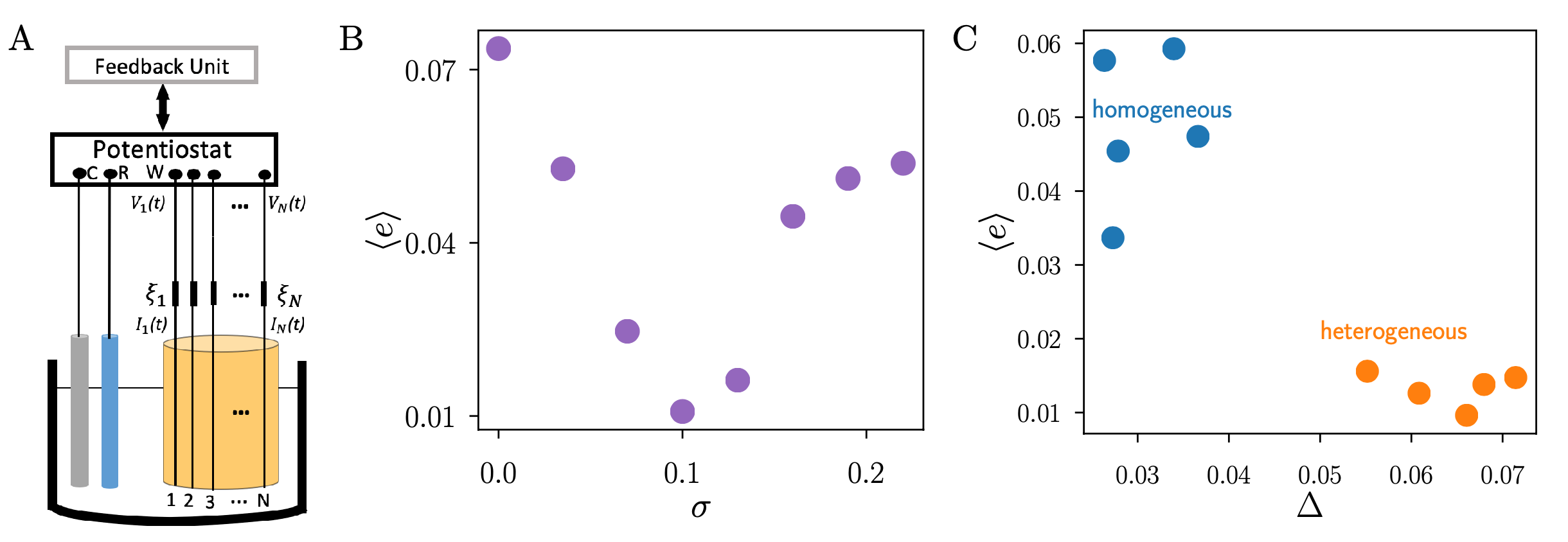}
\caption{Electrochemical oscillator experiments showing that random heterogeneity promotes synchronization. 
({\it A}) Diagram illustrating the setup of the experimental system, where C is the counter electrode, R is the reference electrode, and W are the working electrodes.
({\it B}) Time-averaged synchronization error $\langle e \rangle$ as a function of the nominal oscillator heterogeneity $\sigma$ for one realization of heterogeneous resistances. 
({\it C}) Time-averaged synchronization error $\langle e \rangle$ vs.\ measured oscillator heterogeneity $\Delta$, where each dot represents a different realization of heterogeneous (orange) and homogeneous (blue) systems for $\sigma=0.13$ kohm and $\sigma=0$ kohm, respectively.
}
\label{fig:6}
\end{figure*}

For relatively strong coupling ($K \approx -0.40$ V/mA) and no delay ($\tau = 0$ s), the system exhibits in-phase synchronization \cite{Zhai:2008kv}. 
Similar in-phase synchronization exists for large delay ($\tau \approx 2.4$ s) that corresponds to the mean period of the uncoupled oscillations. 
When $\tau$ is set to $\approx 1.2$ s (about half of the oscillation period), the system exhibits a two-cluster state in which every other element on the grid is in phase, and the neighboring elements are in anti-phase. 
When the delay is set between these two regions ($\tau \approx 1.75 $ s), the system exhibits a desynchronized state. 
{\it Nominal} oscillator heterogeneity was introduced by setting the resistance of each oscillator to a different value $\xi_j$ while keeping the mean resistance fixed to $\overline{\xi}=1.06$ kohm. 
The level of nominal heterogeneity is measured by the standard deviation $\sigma$ among all $\xi_j$.

First, we randomly picked one realization of heterogeneity and experimentally tested its effect on the collective dynamics at different levels of heterogeneity $\sigma$.
Each experiment was initiated close to the in-phase synchronization state and consisted of running the system for $600$ seconds.
The level of coherence was measured by the synchronization error $e(t)$, defined as the standard deviation among the currents $I_j$ at time $t$.
In this case, the synchronization error is a more natural measure of coherence than order parameters because the experimental system is not in the close vicinity of a Hopf bifurcation and the dynamics of the amplitude variables are oscillatory.
(Nevertheless, we verified that the order parameters of the phases extracted using either Hilbert transform or peak detection algorithms give similar results as the ones obtained using the synchronization error.)
The experimental results summarized in \cref{fig:6}{\it B} reveal a well-defined minimum of the average synchronization error $\langle e \rangle$ (averaged over the last $200$ seconds of each experimental run) for an intermediate level of nominal heterogeneity, $\sigma = 0.1$ kohm.
This optimization of synchronization at intermediate levels of heterogeneity is consistent with what we observed numerically for delay-coupled Stuart-Landau oscillators.

Unlike the idealized systems used in simulations, experimental systems come with unavoidable imperfections and uncertainties. 
As a result, the electrochemical oscillators in our experiments have slightly different dynamics even when the resistances are all set to the same nominal value.
These relatively small inherent heterogeneities can arise because of unavoidable differences in the metal wires (e.g., in composition and size) and surface conditions (oxide film layer thickness, localized corrosion, etc.).
To account for such inherent heterogeneity, we use peak detection algorithms \cite{peak} to extract the natural frequency and amplitude of each uncoupled oscillator, and we use that information to calculate the {\it measured} oscillator heterogeneity $\Delta$ for both systems with homogeneous $\xi_j$ and systems with heterogeneous $\xi_j$.
Here, $\Delta = \overline{\rho}_{T} + \overline{\rho}_A$, where $\overline{\rho}_{T}$ ($\overline{\rho}_{A}$) is the standard deviation of the oscillation periods (amplitudes) of the uncoupled oscillators normalized by the mean.
(For additional details on the data analysis protocol, see {\it Materials and Methods}.)
In \cref{fig:6}{\it C}, we show results for five sets of independent experiments.
Each experiment corresponds to a different realization of heterogeneous resistances (for $\sigma$ fixed at $0.13$ kohm), and of the homogeneous system (corresponding to $\sigma=0$ kohm).
It can be seen that when uncoupled, all heterogeneous systems have a much higher measured oscillator heterogeneity $\Delta$ than the homogeneous systems.
In contrast, when coupled, the heterogeneous systems achieve significantly better coherence than the homogeneous systems, which is reflected by a consistently smaller $\langle e \rangle$.

\begin{figure*}[t]
\centering
\includegraphics[width=1\linewidth]{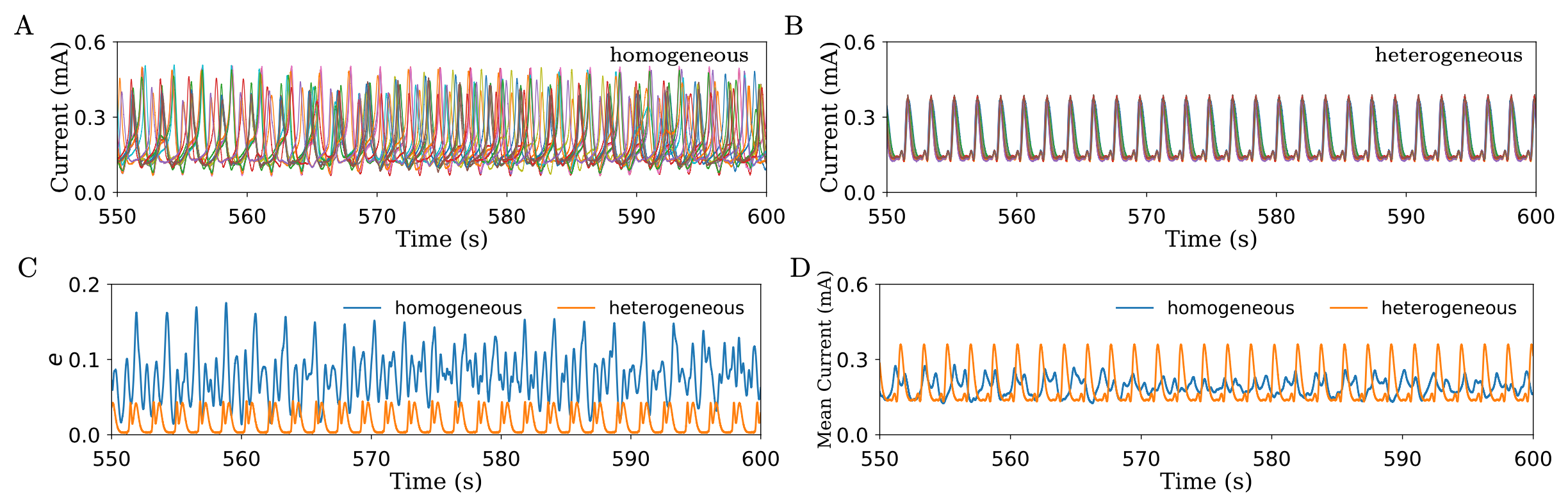}
\vspace{-5mm}
\caption{Comparison between the dynamics of homogeneous and heterogeneous systems in the electrochemical oscillator experiment.
({\it A}) Time series of the currents in the homogeneous system, which shows desynchronized dynamics. 
({\it B}) Time series of the currents in the heterogeneous system, showing that they remain synchronized throughout the experiment. 
({\it C--D}) Evolution of the synchronization error $e$ ({\it C}) and mean-field current ({\it D}) for the two systems. 
In all panels, we show the last $50$ s of trajectories of $600$ s, initialized close to a synchronized state for $\sigma=0$ kohm (homogeneous system) and $\sigma=0.13$ kohm (heterogeneous system).}
\label{fig:7}
\end{figure*}

The striking difference between the behavior of the homogeneous and heterogeneous systems is further visualized in \cref{fig:7}.
There, we compare the dynamics in the first ($\sigma=0$ kohm) and the fifth ($\sigma=0.13$ kohm) data points from \cref{fig:6}{\it B}.
The time series of the homogeneous system (\cref{fig:7}{\it A}) is very much incoherent compared to that of the heterogeneous system (\cref{fig:7}{\it B}).
Accordingly, the heterogeneous system exhibits a smaller synchronization error and a more regular rhythm, as shown in Figs.~\ref{fig:7}{\it C} and {\it D}.
In {\it SI Text} and Fig.~S7, we show that the same phenomenon can be observed when random shortcuts are added to the $4$-by-$4$ lattice (which creates a small-world network \cite{watts1998collective}), suggesting that our conclusions extend to networks with heterogeneous degrees.

\section*{Discussion}

It is often challenging, if not impossible, to completely eliminate component mismatches in oscillator networks. 
Our results suggest that, rather than trying to erase these imperfections (often to no avail), one may instead be able to take advantage of them to promote synchronization required for the system to function. 
Indeed, our theory, simulations, and experiments consistently show that synchronization can often be stabilized by intermediate levels of random oscillator heterogeneity.
The fact that no fine-tuning of the heterogeneity profile is needed to induce synchronization can be valuable for stabilizing synchronization in both technological and biological systems.
For example, it is often important to generate high-power output of coherent light in laser systems.
Semiconductor diode lasers are of interest in many applications due to their low cost, portability, and ease of fabrication, but a single diode laser typically generates an output power of no more than a few watts \cite{ohtsubo2012semiconductor}.
It is thus desirable to couple many diode lasers together and exploit their frequency synchronization to increase the emission power \cite{nair2018phase}.
While in practice no two lasers are perfectly identical, this study indicates that it might be possible to boost the performance of coupled laser arrays by harnessing rather than eliminating the existing mismatches.

In physiology, many important rhythmic processes also depend on the coordination and coherence among a diverse population of cells \cite{glass2001synchronization}.
The heartbeat, for example, is generated by the synchronized activity of thousands of cardiac pacemaker cells in the sinoatrial node \cite{michaels1987mechanisms,maltsev2011synchronization}, whereas the sleep-wake cycle is regulated by the mutual entrainment of circadian cells in the suprachiasmatic nucleus \cite{leloup2003toward,gonze2005spontaneous,to2007molecular}.
Our findings thus raise the question of whether the heterogeneity among pacemaker or circadian cells is a limitation of the biology, or, instead, a feature selected for by evolution to promote synchronization and stabilize vital rhythms in living organisms.
On the other hand, in situations in which synchronization is undesirable, such as epilepsy \cite{jiruska2013synchronization}, the effect demonstrated here can potentially explain why these pathological states appear to be persistent and difficult to suppress despite the inherent diversity of neuronal populations.
This, in turn, might lead to new ideas for therapeutic interventions.

The effect of disorder is also a recurring theme in condensed matter physics \cite{nandkishore2015many}. 
For example, exotic materials such as topological insulators have attracted a vast amount of attention over the past decade \cite{hasan2010colloquium,qi2011topological}.
A defining property of topological insulators is the existence of edge states that are protected by time-reversal symmetry, which makes the states robust to weak disorder.
In the context of oscillator networks, we have been able to go one step further and identify systems and parameter regions for which synchronization is not only immune to disorder, but also enhanced by it.

There are also interesting similarities and differences between the phenomenon described here and noise-induced synchronization \cite{zhou2002noise2,ullner2009noise}.
It is well established that spatially correlated (e.g., common) noise can facilitate synchronization \cite{teramae2004robustness,nakao2007noise,nagai2010noise}, even if the noise is temporally uncorrelated (i.e., white).
In contrast, the disorder we consider here is spatially uncorrelated and temporally quenched.
Understanding how the spatial and temporal features of disorder and noise influence a system's collective dynamics has been an ongoing research effort and a source of new insights.
For example, it has been shown that quenched disorder can induce coherence resonance in driven bistable systems \cite{tessone2006diversity} and that spatially uncorrelated noise can outperform common noise in increasing coherence when oscillators are nonidentical \cite{zack2020coherent}.
Conversely, it has also been shown that quenched disorder can mitigate desynchronization instabilities caused by noise \cite{molnar2020network}.

Finally, it is instructive to reflect on three salient characteristics of the results established here. 
First, oscillator heterogeneity can stabilize frequency synchronization states that are similar to the otherwise unstable states observed in the absence of heterogeneity. 
This is important because it shows that the stability provided by heterogeneity does not come at the price of qualitatively changing the nature of the dynamics.
Second, this stabilization is achieved with high success rate using random parameter heterogeneity, making it easy to implement in real-world systems. 
Third, the effect can be observed for a wide range of network structures, including networks in which all oscillators are identically coupled.
The latter is significant because it shows that the stabilizing effect of oscillator heterogeneity is more fundamental than just counterbalancing instabilities that could have been caused by heterogeneities in the network structure.
Thus, our results reveal an important avenue through which system disorder can give rise to emergent dynamical order.
Future studies further exploring the relation between system disorder and dynamical order will undoubtedly deepen our understanding of collective behavior and of new means to stabilize and control the dynamics of complex systems.

%\appendix

\section*{Materials and Methods} 

\textbf{Numerical Procedure.} Delay-coupled Stuart-Landau oscillators were simulated by employing the {\texttt dde23} integrator in MATLAB, with the relative and absolute tolerances both set to $10^{-4}$. To initialize an oscillator network, we introduce a random perturbation of the order of $10^{-1}$ to the synchronization state at $t=0$. Each system was then evolved for $10^4$ time units, which is long enough for the oscillators to settle into either a coherent state (if synchronization is stable) or an incoherent state (if synchronization is unstable). Our code for simulating delay-coupled Stuart-Landau oscillators can be found at \url{https://github.com/y-z-zhang/disorder_sync}.

\textbf{Experimental Protocols.} The experiments were performed using a standard three-electrode cell with a platinum counter, a \ce{Hg}/\ce{Hg2SO4}/sat.\ce{K2SO4} reference, and a nickel array working electrode. 
The electrolyte was 3M \ce{H2SO4} at \SI{10}{\degree C}.
The electrode array consisted of sixteen \SI{1}{\mm} diameter nickel wires with a spacing of 3 mm. 
The wires were embedded in epoxy, so that only the wire ends were exposed to the electrolyte. 
Before the experiments, the electrode array was polished with a series of sandpapers. 
A multichannel potentiostat (Gill-IK64, ACM Instruments), interfaced with a real-time LabVIEW controller \cite{bitter2017labview}, was used to measure the current $I_j(t)$ and set the potential $V_j(t)$ of the $j$th wire according to Eq.~\ref{eq:exp} at a rate of 200 Hz.
Throughout the experiments we set the circuit potential to $V_0=$ \SI{1.24}{\volt}. 
Without heterogeneity, the individual resistors were set to \SI{1.06} kohm. 
Heterogeneity was introduced by setting the individual resistors to different nominal values drawn from a normal distribution while keeping the mean resistance fixed at \SI{1.06} kohm. 
To avoid accidentally balancing out the inherent heterogeneity, only random realizations of nominal heterogeneity that had a negligible correlation with the natural frequencies of the unperturbed oscillators were used (we required that the absolute value of the correlation coefficient be smaller than $0.2$).
The coupling delay $\tau$ was set to 75\% of the mean natural period of the oscillations, which corresponds to $\tau$ in the range of \SI{1.50}{\second} to \SI{1.75}{\second} throughout the experiments. 
The coupling strength $K$ was set to values about 10\% larger in magnitude than the desynchronization threshold (between $-0.48$ and $-0.40$ V/mA in the reported experiments).

\textbf{Data Analysis Protocols.} The peak detection algorithm finds all local maxima by comparing the neighboring values in a time series.
The mean of the detected peaks is taken as the oscillation amplitude and the mean distance between consecutive peaks is the oscillation period.
Our data analysis scripts and experimental data are available at \url{https://github.com/y-z-zhang/disorder_sync}.
By following the Jupyter Notebooks included in the GitHub repository, one can explore the data interactively and reproduce the results presented in Figs.~\ref{fig:6}, \ref{fig:7} and S7.

\section*{Acknowledgments}
The authors thank Eberhard Bodenschatz, Kyoung-Jin Lee, and Yehuda Braiman for insightful discussions. Y.Z. and A.E.M. were supported by ARO Grants No.\ W911NF-15-1-0272 and No.\ W911NF-19-1-0383. J.L.O.-E. and I.Z.K. were supported by NSF Grant No.\ CHE-1900011. J.L.O.-E. also acknowledges support from CONACYT; and Y.Z. acknowledges support from the Schmidt Science Fellowship.

\bibliography{net_dyn}

\end{document}